\documentclass[a4paper,12pt]{article}

\usepackage[utf8]{inputenc}      
\usepackage[T1]{fontenc}         
\usepackage[english]{babel}      
\usepackage{amsmath,amssymb}     
\usepackage{graphicx}            
\usepackage[nottoc]{tocbibind}   
\usepackage{float}               
\usepackage{booktabs}            
\usepackage{array}               
\usepackage{caption}             
\usepackage{hyperref}            
\usepackage{geometry}            
\usepackage{setspace}            
\usepackage{cite}                

\title{\textbf{Impact of the Pandemic on Currency Circulation in Brazil: Projections using the SARIMA Model}}
\author{%
  \textbf{João Victor Monteiros de Andrade\footnote{Department of Computing, University of Brasília; jotandrade98@gmail.com}}\\[1ex]
  \textbf{Leonardo Santos da Cruz\footnote{Department of Computing, University of Brasília; leonardo-7238@hotmail.com}}
}

\begin{document}

\maketitle

\begin{abstract}

This study analyzes the impact of the COVID-19 pandemic on currency circulation in Brazil by comparing actual data from 2000 to 2023 with counterfactual projections using the \textbf{SARIMA(3,1,1)(3,1,4)\textsubscript{12}} model. The model was selected based on an extensive parameter search, balancing accuracy and simplicity, and validated through the metrics MAPE, RMSE, and AIC. The results indicate a significant deviation between projected and observed values, with an average difference of R\$ 47.57 billion (13.95\%). This suggests that the pandemic, emergency policies, and the introduction of \textit{Pix} had a substantial impact on currency circulation. The robustness of the SARIMA model was confirmed, effectively capturing historical trends and seasonality, though findings emphasize the importance of considering exogenous variables, such as interest rates and macroeconomic policies, in future analyses. Future research should explore multivariate models incorporating economic indicators, long-term analysis of post-pandemic currency circulation trends, and studies on public cash-holding behavior. The results reinforce the need for continuous monitoring and econometric modeling to support decision-making in uncertain economic contexts.

\textbf{Keywords:} Currency circulation, SARIMA model, COVID-19 pandemic, monetary policy, econometric modeling.

\end{abstract}

\section{Introduction}
\subsection{Contextualization}
\subsubsection{The importance of currency circulation for the economy}

\quad \quad \, Currency circulation is a fundamental element for the functioning of any economy. It is directly related to the volume of money that flows between economic agents—companies, consumers, the government, and financial institutions—over a given period. This flow is essential for enabling commercial transactions, investments, salary payments, and the consumption of goods and services.

In an economy, money fulfills several functions, such as:
\begin{enumerate}
    \item \textbf{Medium of exchange}: Facilitates transactions by eliminating the inefficiencies of bartering;
    \item \textbf{Unit of account}: Serves as a standard for measuring the value of goods and services;
    \item \textbf{Store of value}: Allows purchasing power to be preserved over time.
\end{enumerate}

The amount of money in circulation directly influences economic indicators such as inflation, consumption levels, economic activity, and monetary policy. For example, an excessive increase in money circulation can lead to inflationary pressures, while a shortage can harm consumption and investment dynamics. Therefore, monetary authorities, such as the Central Bank of Brazil (BACEN), monitor and control the money flow to maintain economic stability.

\subsubsection{The economic impact of the pandemic in Brazil}

\quad \quad \, The COVID-19 pandemic, which began in March 2020, caused an unprecedented global health crisis that quickly turned into an economic crisis. Brazil was no exception and faced severe impacts in various dimensions:

\begin{enumerate}
    \item \textbf{Decline in economic activity}: Social isolation and government-imposed restrictions to contain the spread of the virus led to the shutdown of key economic sectors such as commerce, services, and industry, causing a drop in Brazil's Gross Domestic Product (GDP) in 2020;
    \item \textbf{Impact on the labor market}: There was a significant increase in unemployment and a decline in household income, reducing purchasing power and consumption;
    \item \textbf{Economic stimulus policies}: To mitigate the economic effects, the federal government implemented emergency measures such as the Emergency Aid program, which injected billions of reais directly into the economy, increasing currency circulation. This program was crucial in sustaining the consumption of low-income families during the crisis \cite{muniz2020impactos};
    \item \textbf{Increase in the monetary base}: The need for liquidity to finance emergency policies, along with market uncertainty, led to a significant rise in the amount of money in circulation. This increase brought additional challenges for monetary policy, such as the risk of inflationary pressures and exchange rate imbalances;
    \item \textbf{Changes in consumption patterns and money usage}: During the pandemic, there was a rise in demand for physical cash, possibly due to economic uncertainty and concerns about access to banking services. Additionally, the crisis accelerated financial digitalization, with a higher use of electronic payments and instant transfers, such as \textit{Pix} \cite{amboage2024technological} \cite{duarte2022central}.
\end{enumerate}

\subsection{Research Problem}

\quad \quad \,\, \emph{What would currency circulation in Brazil be like under normal circumstances, without the effects of the pandemic?}

In the years before the pandemic, time series data on currency circulation exhibited clear patterns, such as trends (long-term movements) and seasonality (periodic fluctuations). These patterns were stable, reflecting consistent economic behaviors over time.

With the arrival of the pandemic, these regularities were abruptly altered due to economic changes. Thus, projecting a counterfactual scenario allows us to understand how these patterns might have remained unchanged in the absence of the shocks caused by the health crisis.

\subsection{Objectives}

\quad \quad \, The objective of this study is to project Brazil’s currency circulation in a scenario without the effects of the pandemic, using pre-pandemic time series data. To achieve this, the historical series will be modeled with the \textbf{SARIMA} model, allowing for a counterfactual projection that represents the expected behavior of currency circulation under normal conditions. Subsequently, the projected results will be compared with actual values observed during the pandemic to identify and quantify the crisis's impact on this economic variable.

\subsection{Justification}

\quad \quad \, This study is justified by the need to understand the effects of the COVID-19 pandemic on currency circulation, a crucial variable for the functioning of the Brazilian economy. The crisis brought substantial changes in economic behavior, including emergency policies, increased liquidity, and altered consumption patterns, directly affecting the amount of money in circulation.

By projecting a counterfactual scenario, it is possible to:
\begin{itemize}
    \item Assess the real impact of the pandemic on currency circulation;
    \item Understand how the crisis altered previously observed economic patterns;
    \item Contribute to the development of more effective economic policies, assisting the Central Bank and other institutions in planning future actions in response to similar events.
\end{itemize}

Additionally, the choice of classical models such as SARIMA is justified by their effectiveness in capturing historical patterns with low mathematical complexity and reduced computational cost, ensuring an accessible interpretation of results.

\section{Literature Review}

\quad \quad \, Currency circulation refers to the volume of money in transactions and available within the economic system over a given period. It can be analyzed through monetary aggregates, which represent different levels of money liquidity \cite{mishkin2015economics}:
\begin{itemize}
    \item \textbf{M1}: Consists of paper money held by the public and demand deposits (high liquidity);
    \item \textbf{M2}: Includes M1 plus savings deposits and short-term securities;
    \item \textbf{M4}: Encompasses M2 and less liquid financial instruments, such as public and private bonds.
\end{itemize}

Currency circulation is closely linked to classical economic theories, such as the \textbf{Quantity Theory of Money}, formulated by Irving Fisher \cite{fisher1920}. According to this theory, the volume of money in circulation ($M$) is related to the price level ($P$), the volume of transactions ($T$), and the velocity of money circulation ($V$), expressed by the equation:
\[
M \cdot V = P \cdot T.
\]

This relationship indicates that changes in $M$ can directly affect prices and economic activity, especially in times of monetary expansion or liquidity restriction \cite{mankiw2007macroeconomics}. 

\subsection{Economic Impacts of the Pandemic}

\quad \quad \, Some studies address the economic impacts of the pandemic on variables such as GDP, unemployment, and digital financial transactions, but relatively few works specifically investigate the dynamics of currency circulation in Brazil using advanced statistical techniques. Internationally, some studies discuss changes in monetary circulation patterns in other countries, but without direct application to the Brazilian case. 

In Brazil, data from the Central Bank indicate that the amount of money in circulation has grown substantially since 2019, even with the advancement of payment technologies such as \textit{Pix}. Nevertheless, national literature lacks studies that statistically model the direct and indirect impacts of the pandemic on this dynamic. While changes in the use of banknotes and physical coins are often attributed to factors such as economic uncertainties and liquidity demand, there is no consensus supported by methodologically robust analyses. Thus, examining how the pandemic and policies like \textit{Pix} affected currency circulation is crucial to understanding this atypical period.

\subsection{SARIMA Model: Definition and Components}

\quad \quad \, The SARIMA (Seasonal AutoRegressive Integrated Moving Average) model is an extension of the ARIMA model that includes seasonal components to handle time series exhibiting periodic patterns. It is represented by the set of parameters $(p,d,q)\times (P,D,Q)_{s}$, where:
\begin{itemize}
    \item $(p,d,q)$: non-seasonal components (orders of the autoregressive part, differencing, and moving average);
    \item $(P,D,Q)$: seasonal components (orders of the autoregressive, differencing, and moving average parts of seasonality);
    \item $s$: seasonal period (e.g., $s=12$ for monthly data).
\end{itemize}

Mathematically, it can be written as:
\[
\Phi_P(B^s)\,\phi_p(B)\,(1 - B)^d\,(1 - B^s)^D X_t \;=\; \Theta_Q(B^s)\,\theta_q(B)\,\varepsilon_t,
\]
where $B$ is the lag operator, $\phi_p(B)$ and $\theta_q(B)$ are polynomials of orders $p$ and $q$, $\Phi_P(B^s)$ and $\Theta_Q(B^s)$ handle the seasonal components, $(1-B)^d$ and $(1-B^s)^D$ represent the non-seasonal and seasonal differencing, and $\varepsilon_t$ is white noise \cite{box2015time,hamilton1994time}.

\subsubsection{Assumptions About Residuals}

\quad \quad \, For the SARIMA model to be appropriate, the residuals ($\varepsilon_t$) must meet certain assumptions:
\begin{enumerate}
    \item \textbf{Zero mean}: $E(\varepsilon_t) \approx 0$, indicating no systematic bias;
    \item \textbf{No autocorrelation}: $\text{Corr}(\varepsilon_t,\varepsilon_{t-k}) \approx 0$, for any $k \neq 0$;
    \item \textbf{Normality}: Residual distribution should be approximately normal, relevant for inference and forecasting interval construction.
\end{enumerate}

These assumptions can be verified through statistical tests such as \textit{Box-Pierce} (autocorrelation)\cite{boxpierce}, \textit{Dickey-Fuller} (stationarity) \cite{adf}, and \textit{Kolmogorov-Smirnov} (normality)\cite{ks},\cite{smirnov}, as well as graphical inspections \cite{hyndman2018forecasting}.

\subsubsection{Economic Applications}

\quad \quad \, SARIMA is widely used in various economic applications, including \emph{money demand}, electricity consumption, agricultural price forecasting, and more \cite{box2015time}. Its ability to capture both trend and seasonal components makes it particularly useful in scenarios where data exhibit regular fluctuations.

\section{Methodology}

\subsection{Currency Circulation (2000--2023)}

\quad \quad \, The data used in this study were obtained from the Central Bank of Brazil (BACEN) database. They consist of time series representing the volume of currency circulation in Brazilian reais (R\$) from January 2000 to May 2023.

\begin{figure}[H]
    \centering
    \includegraphics[width=1\textwidth]{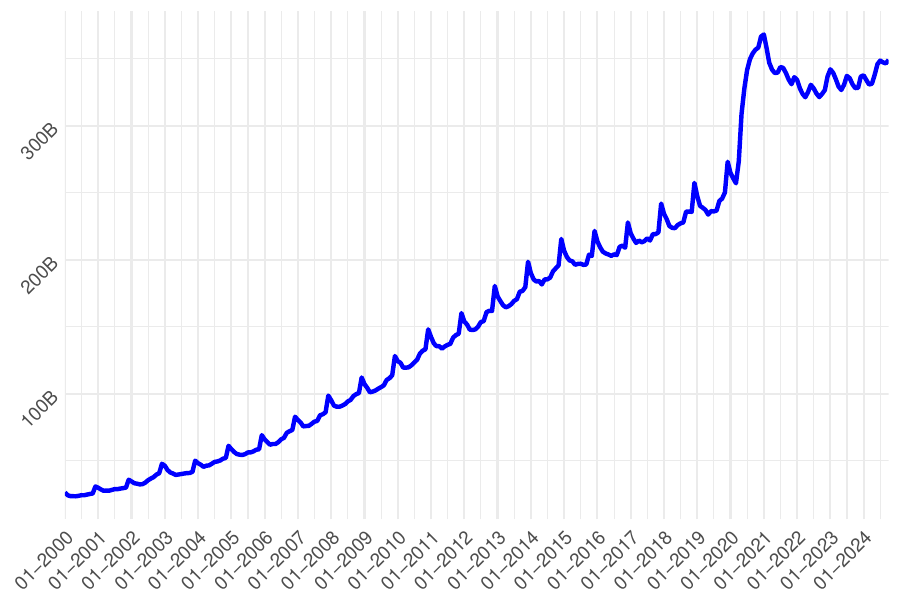} 
    \caption{Time series of currency circulation in reais in Brazil (2000--2024).}
    \label{fig:serie_circulacao}
\end{figure}

For analysis purposes, the series was divided into three distinct parts:
\begin{itemize}
    \item \textbf{Training set}: Covers the period from January 2000 to March 2020, totaling 231 values. This interval is used for training the SARIMA model, aiming to adjust its parameters to the historical patterns of the series;
    \item \textbf{Test set}: Includes the subsequent 12 months (from April 2020 to March 2021), used to validate the model by evaluating its performance with metrics such as MSE (Mean Squared Error), MAPE (Mean Absolute Percentage Error), and RMSE (Root Mean Squared Error);
    \item \textbf{Comparison set}: Consists of 37 values (from April 2021 to May 2022), intended to analyze the behavior of currency circulation during the officially recognized pandemic period.
\end{itemize}

\subsection{Justification for the SARIMA Model}

\quad \quad \, The choice of the \textbf{SARIMA} (Seasonal Autoregressive Integrated Moving Average) model is due to its ability to model time series with seasonal and non-stationary components, which are common in economic variables affected by periodic fluctuations. Compared to simpler models (such as moving averages or linear regression), SARIMA more robustly incorporates historical autocorrelations and seasonal patterns. Furthermore, it offers relatively accessible interpretability for economists and policymakers, along with moderate computational cost. 

\subsection{Methodological Limitations}

\quad \quad \, Despite its advantages, using SARIMA also presents certain limitations:
\begin{itemize}
    \item \textbf{Data quality}: Gaps, outliers, or inconsistencies can affect model fitting;
    \item \textbf{Unexpected structural changes}: Shocks such as financial crises, pandemics, or radical monetary policies may alter historical patterns;
    \item \textbf{Univariate nature}: The model does not account for exogenous variables (such as GDP, interest rate, inflation), which may limit the explanation of certain shocks;
    \item \textbf{Forecasting range}: SARIMA is generally more suitable for short- to medium-term projections, and its accuracy may decrease for longer horizons.
\end{itemize}

\subsection{Estimation of Model Parameters}

\quad \quad \, To estimate the parameters of the SARIMA model, an exhaustive search was conducted, considering various orders of seasonal and non-seasonal components within the following ranges:
\[
p, q, P, Q \in [0, 5], \quad d, D \in [0, 1], \quad s = 12.
\]

The values of \(d\) and \(D\) were restricted to \([0,1]\) based on preliminary tests, which demonstrated that applying a single differencing (seasonal and non-seasonal) was sufficient to achieve stationarity in the series while avoiding over-differencing that could distort its structure.

The initial evaluation of each combination was performed using the \textbf{Mean Squared Error (MSE)} on the test window. Then, the 10 models with the lowest MSE underwent a complementary analysis using the following metrics: Root Mean Squared Error (RMSE), Mean Absolute Percentage Error (MAPE), and the Akaike Information Criterion (AIC), calculated on the training window.

The combined use of these metrics allowed for a balanced assessment of modeling accuracy, forecasting quality, model simplicity, and proper data fitting. Consequently, the model that best reconciled low forecast errors (RMSE and MAPE) with the lowest AIC was selected.

\section{Analysis and Discussion of Results}

\quad \quad \, After applying the described procedure, several SARIMA models were estimated. Table~\ref{tab:parametros_sarima} presents the tested parameter combinations and their associated performance metrics: \emph{RMSE}, \emph{MAPE}, and \emph{AIC}.

\begin{table}[H]
\centering
\caption{Estimated parameters of SARIMA models with their respective errors.}
\label{tab:parametros_sarima}
\begin{tabular}{cccccccccc}
\toprule
$p$ & $d$ & $q$ & $P$ & $D$ & $Q$ & \textbf{RMSE} & \textbf{MAPE} & \textbf{AIC} \\
\midrule
5 & 1 & 3 & 2 & 1 & 3 & 1077009 & 0.72\% & 6770.556 \\
3 & 1 & 0 & 3 & 1 & 5 & 1129713 & 0.73\% & 6770.362 \\
3 & 1 & 1 & 3 & 1 & 4 & 1121844 & 0.72\% & 6770.030 \\
1 & 1 & 4 & 3 & 1 & 4 & 1117261 & 0.72\% & 6770.706 \\
3 & 1 & 0 & 3 & 1 & 4 & 1119026 & 0.72\% & 6768.438 \\
1 & 1 & 2 & 3 & 1 & 5 & 1138818 & 0.73\% & 6771.089 \\
5 & 1 & 2 & 2 & 1 & 3 & 1124421 & 0.74\% & 6770.437 \\
5 & 1 & 1 & 3 & 1 & 5 & 1116131 & 0.72\% & 6773.843 \\
0 & 1 & 3 & 3 & 1 & 5 & 1138901 & 0.73\% & 6768.707 \\
5 & 0 & 1 & 3 & 1 & 2 & 1147468 & 0.75\% & 6810.198 \\
\bottomrule
\end{tabular}
\end{table}

It is observed that the first model (\(p=5,\,d=1,\,q=3,\,P=2,\,D=1,\,Q=3\)) has the lowest \textbf{RMSE} (1,077,009), which at first glance might suggest good absolute performance. However, when examining the other metrics, it becomes evident that there are alternative models that offer a better balance between relative error and complexity.

Among the analyzed options, the \textbf{third model} (\(p=3,\,d=1,\,q=1,\,P=3,\,D=1,\,Q=4\)) stood out for the following reasons:

\begin{itemize}
    \item \textbf{Lowest AIC} (6770.030): This value suggests a better balance between model fit quality and complexity, avoiding overfitting.
    \item \textbf{Lowest MAPE} (0.72\%): The lowest mean absolute percentage error indicates that, proportionally, the forecasts are more consistent with the actual values, an important factor for series with large magnitudes.
\end{itemize}

Although its RMSE is slightly higher than the best value found, its superior performance in percentage error (MAPE) and lower AIC make the \textbf{SARIMA(3,1,1)(3,1,4)\textsubscript{12}} model preferable. In cases where the absolute value of the series is very high, proportional metrics such as MAPE tend to be more meaningful for decision-making. Additionally, the complexity penalty captured by AIC reinforces the robustness of this choice.

Thus, the \textbf{SARIMA(3,1,1)(3,1,4)\textsubscript{12}} model proved to be the best solution, as it adequately balances series fitting, model simplicity, and predictive quality. This selection highlights the importance of using multiple metrics to evaluate time series models, especially when dealing with highly variable data, where relative error differences can be more relevant than absolute fluctuations.

\begin{figure}[H]
    \centering
    \includegraphics[width=1\textwidth]{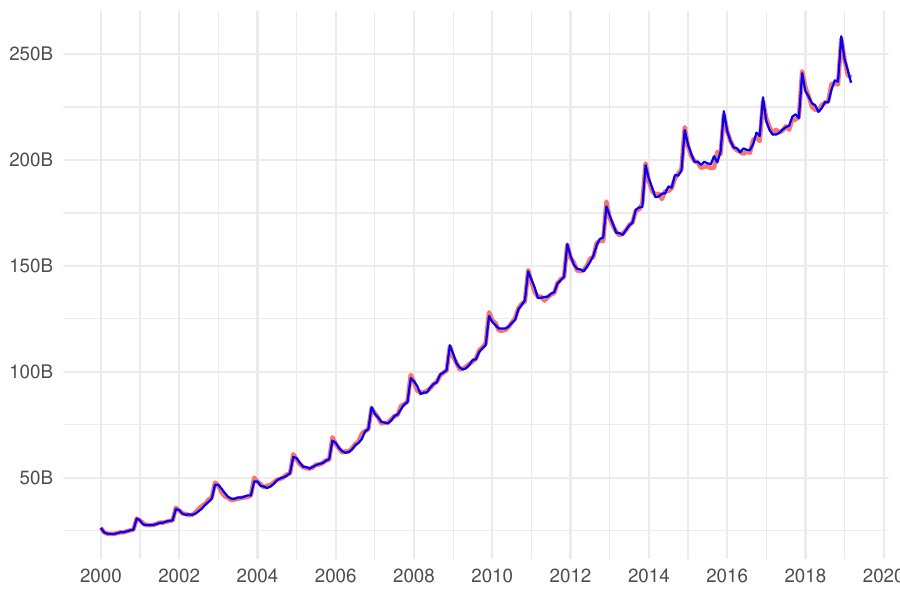}
    \caption{Time series (2000--2020) and SARIMA(3,1,1)(3,1,4)\textsubscript{12} fit (training window).}
    \label{fig:serie_ajuste_sarima}
\end{figure}

\subsection{Performance in the Test Window}

\quad \quad \, Table~\ref{tab:desempenho_teste} presents the performance of different models in forecasting the 12 months (April 2020 to March 2021) immediately following the training period. This one-year window is particularly challenging as it encompasses a complete seasonal cycle (capturing annual trends) and coincides with abrupt economic changes resulting from the onset of the pandemic. Despite these challenges, the \textbf{SARIMA(3,1,1)(3,1,4)\textsubscript{12}} model achieved competitive \emph{RMSE} and \emph{MAPE} values compared to other alternatives, demonstrating strong resilience in handling uncertainties and unpredictable fluctuations during this period.

Given the large magnitude of the series, the use of \emph{MAPE} plays a crucial role in assessing predictive performance, as it provides a relative perspective on error. Even small percentage differences can translate into significant absolute amounts, making the model’s strong performance in this metric even more relevant. Thus, by combining good results in \emph{RMSE} and \emph{MAPE}, the \textbf{SARIMA(3,1,1)(3,1,4)\textsubscript{12}} model not only captures seasonal and trend patterns effectively but also maintains reliable forecasts in a highly volatile scenario. This evidence set makes the model particularly valuable for short-term projections, supporting decision-making in an unstable economic context.

\begin{table}[H]
\centering
\caption{Model performance in the test window}
\label{tab:desempenho_teste}
\begin{tabular}{lcc}
\toprule
\textbf{Model} & \textbf{RMSE} & \textbf{MAPE} \\
\midrule
SARIMA(5,1,3)(2,1,3)\textsubscript{12} & 2803660 & 1.04\% \\
SARIMA(3,1,0)(3,1,5)\textsubscript{12} & 2929427 & 1.07\% \\
SARIMA(3,1,1)(3,1,4)\textsubscript{12} & 2929818 & 1.07\% \\
SARIMA(1,1,4)(3,1,4)\textsubscript{12} & 2943441 & 1.09\% \\
SARIMA(3,1,0)(3,1,4)\textsubscript{12} & 2946947 & 1.08\% \\
SARIMA(1,1,2)(3,1,5)\textsubscript{12} & 2952918 & 1.07\% \\
SARIMA(5,1,2)(2,1,3)\textsubscript{12} & 2962866 & 1.10\% \\
SARIMA(5,1,1)(3,1,5)\textsubscript{12} & 2976435 & 1.09\% \\
SARIMA(0,1,3)(3,1,5)\textsubscript{12} & 3009725 & 1.09\% \\
SARIMA(0,1,3)(3,1,5)\textsubscript{12} & 3035363 & 1.12\% \\
\bottomrule
\end{tabular}
\end{table}

\begin{figure}[H]
    \centering
    \includegraphics[width=.95\textwidth]{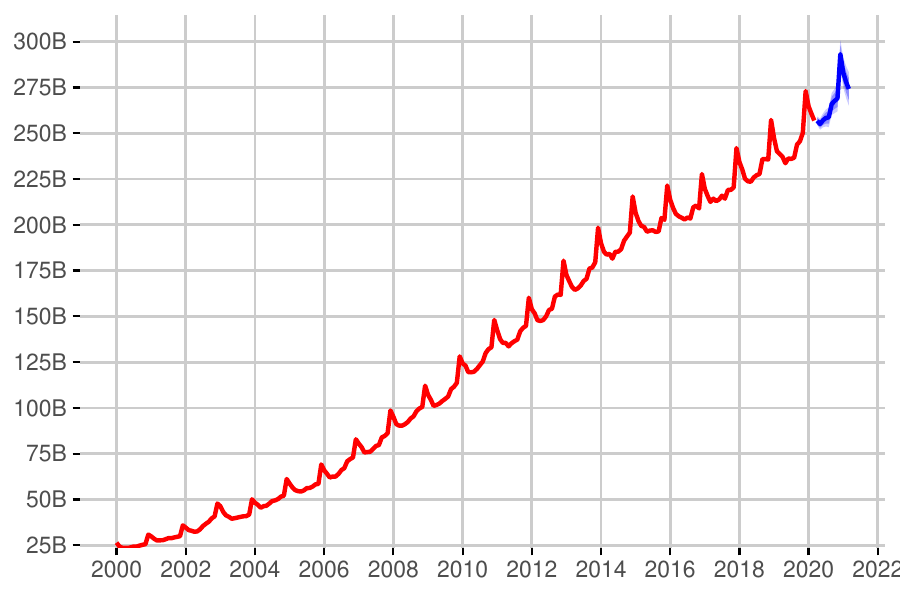}
    \caption{Forecasts for the test window (Apr/2020 to Mar/2021), with confidence intervals.}
    \label{fig:serie_pre_teste}
\end{figure}

\subsection{Verification of Residual Assumptions}

\quad \quad \, To corroborate the adequacy of the \textbf{SARIMA(3,1,1)(3,1,4)\textsubscript{12}} model, statistical tests were performed on the residuals, with the following results:

\begin{enumerate}
    \item \textbf{Zero Mean (ADF Test)}:
    \[
    \text{Dickey-Fuller} = -5.9968, 
    \quad p\text{-value} = 0.01.
    \]
    Since $p < 0.05$, the null hypothesis of non-stationarity is rejected, indicating that the residuals do not exhibit a systematic trend and can be considered stationary.

    \item \textbf{Autocorrelation (Box-Pierce Test)}:
    \[
    \chi^2 = 29.106, \quad p\text{-value} = 0.0857.
    \]
    The hypothesis of no autocorrelation is not rejected \((p > 0.05)\), confirming that the observations are independent over time. (The Box-Ljung test provided a similar result, with \(p \approx 0.051\) \cite{boxjung}.) This is also confirmed by the ACF and PACF plots.

    \begin{figure}[H]
        \centering
        \includegraphics[width=1\textwidth]{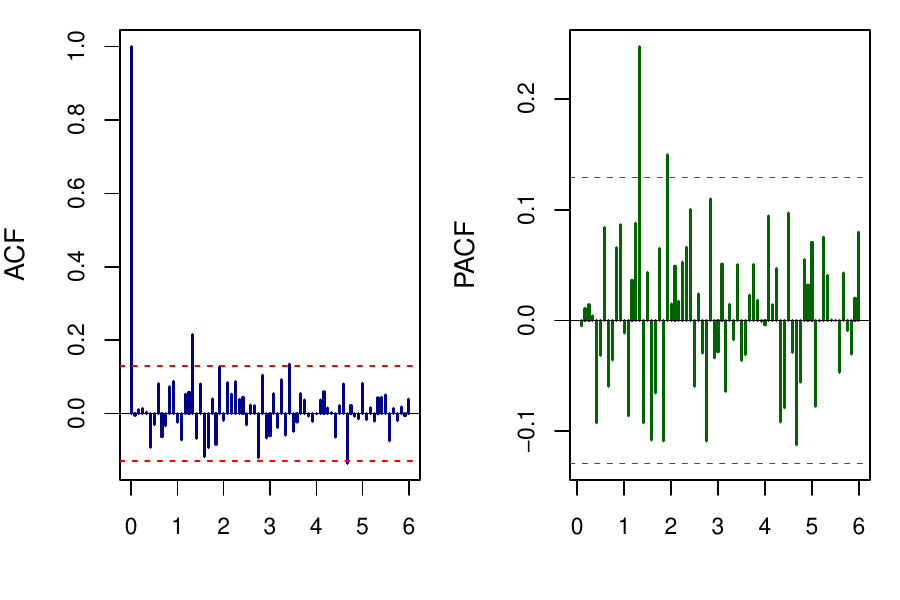}
        \caption{ACF and PACF plots of the time series.}
        \label{fig:time_series}
    \end{figure}

    \item \textbf{Normality (Kolmogorov-Smirnov Test)}:
    \[
    D = 0.084823,
    \quad p\text{-value} = 0.07201.
    \]
    The hypothesis of normality is not rejected $(p>0.05)$, meaning that the residuals can be considered approximately normally distributed.
\end{enumerate}

These results indicate that the \textbf{SARIMA(3,1,1)(3,1,4)\textsubscript{12}} model meets the main adequacy assumptions regarding the residuals: there is no systematic bias, no remaining autocorrelation, and the normality assumption is not violated.

\section{Quantitative Analysis of Economic Impact}

\quad \quad \, To quantify the economic impact of shocks caused by the pandemic and the adoption of \textit{Pix}, the \textbf{real} observed values (\(R_t\)) were compared with the \textbf{predicted} values (\(\hat{R}_t\)) from the SARIMA model, particularly in the period from April 2021 to May 2022. It is defined as follows:

\begin{itemize}
    \item \textbf{Nominal Difference}:
    \[
    I_t^{\text{Nominal}} = R_t - \hat{R}_t,
    \]
    which reflects the absolute deviation in monetary units (R\$);
    \item \textbf{Percentage Difference}:
    \[
    I_t^{\text{Percentual}} = \left(\frac{R_t - \hat{R}_t}{\hat{R}_t}\right) \times 100,
    \]
    which expresses this deviation in relative terms (\%).
\end{itemize}

\subsection{Confidence Intervals (80\% and 95\%)}

\quad \quad \, In addition to point estimates, it is essential to assess the uncertainty range associated with projections, particularly in scenarios with potential external fluctuations beyond historical patterns. In this regard, the lower (\(\hat{R}_{t,\alpha}^{\text{Lower}}\)) and upper (\(\hat{R}_{t,\alpha}^{\text{Upper}}\)) limits of the 80\% and 95\% confidence intervals are considered. These bands reflect probable variations around the central forecast, allowing for a more comprehensive risk analysis. To quantify the deviation both in nominal and relative terms, the following definitions are established \cite{box2015time} \cite{hyndman2018forecasting}:

\[
I_{t,\alpha}^{\text{Nominal,Lower}} = R_t - \hat{R}_{t,\alpha}^{\text{Lower}}, \quad
I_{t,\alpha}^{\text{Nominal,Upper}} = R_t - \hat{R}_{t,\alpha}^{\text{Upper}},
\]
\[
I_{t,\alpha}^{\text{Percentual,Lower}} = \left(\frac{R_t - \hat{R}_{t,\alpha}^{\text{Lower}}}{\hat{R}_{t,\alpha}^{\text{Lower}}}\right) \times 100, \quad
I_{t,\alpha}^{\text{Percentual,Upper}} = \left(\frac{R_t - \hat{R}_{t,\alpha}^{\text{Upper}}}{\hat{R}_{t,\alpha}^{\text{Upper}}}\right) \times 100.
\]

\noindent
\quad \quad \, Thus, both the absolute monetary magnitude of the differences and their proportional impact are examined within the confidence bands, enabling managers and analysts to identify less probable yet possible scenarios of more significant discrepancies.

\subsection{Results of Nominal and Percentage Differences}

\begin{figure}[H]
    \centering
    \includegraphics[width=1\textwidth]{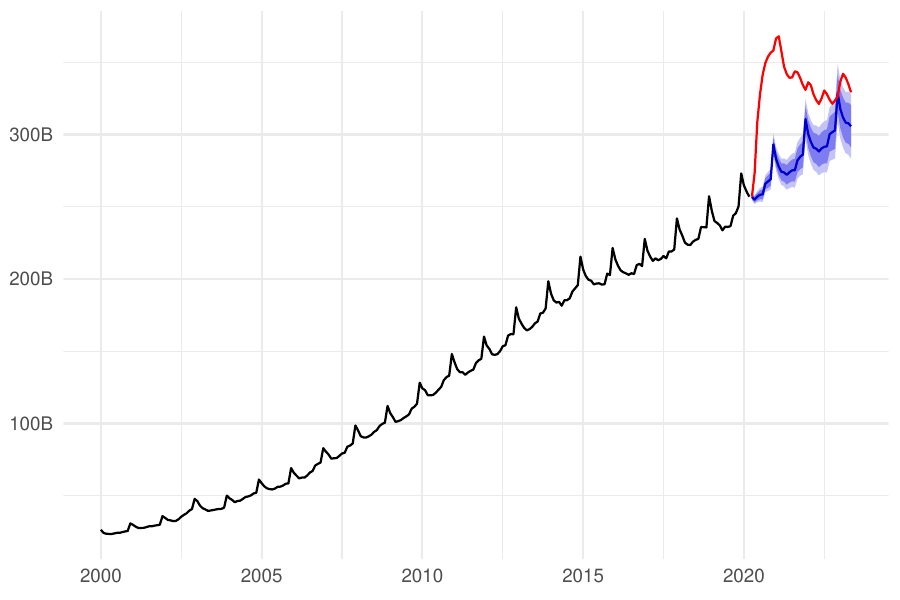}
    \caption{Comparison between actual (black and red) and predicted (blue) time series with 95\% and 80\% confidence interval.}
    \label{fig:serie_comparacao}
\end{figure}

\quad \quad \, To further understand the deviation between actual and forecasted values, both the nominal difference (in reais) and the relative difference (in percentage) were analyzed. This dual approach allows for evaluating both the absolute magnitude of the deviation and its proportional impact on the expected value. Table~\ref{tab:media_diferencas} summarizes these measures, including two confidence intervals (80\% and 95\%) to provide a broader uncertainty range and highlight possible extreme variations. Subsequently, Figure~\ref{fig:serie_comparacao} graphically presents the comparison between the observed and forecasted time series, with emphasis on the period following the onset of the pandemic.

\begin{table}[H]
\centering
\resizebox{\textwidth}{!}{%
  \begin{tabular}{ccc}
    \toprule
    \textbf{Type of Difference} & \textbf{Nominal Difference (R\$)} & \textbf{Percentage Difference (\%)} \\
    \midrule
    \textbf{Point Estimate (Mean)} & 47,567,811,705 & 13.95\% \\
    \midrule
    \multicolumn{3}{c}{\textbf{80\% Confidence Interval}} \\
    Lower Limit & 56,077,421,027 & 16.50\% \\
    Upper Limit & 39,058,202,383 & 11.41\% \\
    \midrule
    \multicolumn{3}{c}{\textbf{95\% Confidence Interval}} \\
    Lower Limit & 60,582,136,093 & 17.85\% \\
    Upper Limit & 34,553,487,317 & 10.06\% \\
    \bottomrule
  \end{tabular}
}
\caption{Mean Nominal and Percentage Differences between Actual and Forecasted Values.}
\label{tab:media_diferencas}
\end{table}

\noindent The following observations can be made:
\begin{itemize}
    \item An \textbf{average nominal difference} of approximately R\$ 47.57 billion, indicating a significant deviation between actual values and expected behavior under normal conditions;
    \item An \textbf{average percentage difference} of 13.95\%, reinforcing that, in relative terms, the impact was substantial;
    \item Wider confidence intervals (95\%) indicate even greater discrepancies in less probable but possible scenarios, reaching up to 17.85\% average difference.
\end{itemize}

These discrepancies support the hypothesis that the pandemic—combined with emergency policies, innovations such as \textit{Pix}, and changes in cash-holding behavior—drove currency circulation beyond what would have been projected in a typical scenario.

\section{Conclusion}

\quad \quad \, The analysis conducted allowed for a comparison between the actual observed values of currency circulation in Brazil from 2000 to 2023 and the projections of a counterfactual scenario free from the shocks of the COVID-19 pandemic. The \textbf{SARIMA(3,1,1)(3,1,4)\textsubscript{12}} model was selected after a careful parameter search, balancing low complexity and high accuracy. It was validated by metrics such as MAPE, RMSE, and AIC and met the assumptions regarding residuals (zero mean, absence of autocorrelation, and normality).

The results highlight that:
\begin{itemize}
    \item \textbf{Significant deviation}: The differences between actual and predicted values (an average of R\$ 47.57 billion or 13.95\%) indicate that the pandemic, emergency policies, and the introduction of \textit{Pix} had a substantial impact on monetary circulation;
    \item \textbf{SARIMA model robustness}: The model effectively captured historical trends and seasonality, performing well in the test window (pre-pandemic);
    \item \textbf{Importance of exogenous variables}: External factors beyond the univariate model, such as interest rates, fiscal policy, and macroeconomic uncertainty, may be crucial in explaining deviations during crisis periods.
\end{itemize}

\noindent As future research directions, we suggest:
\begin{itemize}
    \item \textbf{Multivariate models}: Investigate the inclusion of exogenous variables (e.g., GDP, SELIC, inflation) to refine the understanding of factors influencing currency circulation;
    \item \textbf{Long-term analyses}: Assess whether, after the end of emergency programs, monetary circulation returns to previous patterns or establishes a new level;
    \item \textbf{Studies on cash demand}: Examine cash retention alongside the growth of digital payment methods to capture potential structural changes in public behavior.
\end{itemize}

\noindent In summary, this study indicates that the health crisis and associated policies significantly altered currency circulation, emphasizing the need for continuous analyses and robust econometric models to support decision-making in uncertain scenarios.

\section*{Acknowledgements}

\quad \quad \, The authors express their deep gratitude to CAPES, whose generous funding has been essential to the development of this research, carried out while pursuing their doctoral studies. They thank the University of Brasília, especially for the Time Series Analysis course taught by Professor José Augusto Fiorucci, which enriched their knowledge and guided their studies. João Victor M. de Andrade thanks the Central Bank of Brazil, through DEPEP, whose research grant provided the indispensable environment and support for the progress of this investigation, as well as the staff members Eurilton Alves Araújo Júnior, Alexandre Kornelius, and Angelo Marsiglia Fasolo, whose support was fundamental to the advancement of this work.


\clearpage
\appendix
\section{Appendix}

\quad \quad \, In this section, we present a detailed description of the results obtained for the SARIMA. The information is organized in tables to facilitate the understanding of the estimated parameters and the model performance measures.

\subsection{Model and Specification}
\quad \quad \, The fitted model is an $\boldsymbol{SARIMA(3,1,1)(3,1,4)_{12}}$, which indicates:

\begin{itemize}
    \item \textbf{Non-seasonal part}:
    \begin{itemize}
        \item Autoregressive (AR) order: 3 terms;
        \item Differencing order: 1;
        \item Moving average (MA) order: 1 term.
    \end{itemize}
    \item \textbf{Seasonal part} (with a period of 12 periods, typically months):
    \begin{itemize}
        \item Seasonal autoregressive (SAR) order: 3 terms;
        \item Seasonal differencing order: 1;
        \item Seasonal moving average (SMA) order: 4 terms.
    \end{itemize}
\end{itemize}

This structure shows that the behavior of the series is influenced by immediate (non-seasonal) effects as well as by effects that repeat every 12 periods (seasonal).

\subsection{Estimated Coefficients and Standard Errors}

\quad \quad \, Tables~\ref{tab:coefficients_noseason} and \ref{tab:coefficients_season} present the estimated coefficients for the non-seasonal and seasonal components of the model, along with their corresponding standard errors.

\begin{table}[H]
\centering
\caption{Estimated Coefficients and Standard Errors of the SARIMA Model (Non-seasonal Part)}
\label{tab:coefficients_noseason}
\begin{tabular}{lcc}
\toprule
\textbf{Parameter} & \textbf{Estimated Value} & \textbf{Standard Error} \\
\midrule
ar1    & 0.0225   & 0.2758 \\
ar2    & 0.0991   & 0.0828 \\
ar3    & 0.1577   & 0.0727 \\
ma1    & -0.1814  & 0.2742 \\
\bottomrule
\end{tabular}
\end{table}

\begin{table}[H]
\centering
\caption{Estimated Coefficients and Standard Errors of the SARIMA Model (Seasonal Part)}
\label{tab:coefficients_season}
\begin{tabular}{lcc}
\toprule
\textbf{Parameter} & \textbf{Estimated Value} & \textbf{Standard Error} \\
\midrule
sar1   & -0.6787  & 0.2360 \\
sar2   & 0.9196   & 0.1034 \\
sar3   & 0.7531   & 0.2706 \\
sma1   & 0.0242   & 0.5518 \\
sma2   & -1.4808  & 0.7837 \\
sma3   & -0.1795  & 0.2701 \\
sma4   & 0.6910   & 0.4735 \\
\bottomrule
\end{tabular}
\end{table}

\subsection{Model Performance Measures}

\quad \quad \, Table~\ref{tab:fit} summarizes the main performance indicators and the quality of the model fit.

\begin{table}[H]
\centering
\caption{Model Performance Measures of the SARIMA Model}
\label{tab:fit}
\begin{tabular}{lc}
\toprule
\textbf{Indicator} & \textbf{Value} \\
\midrule
Residual Variance ($\sigma^2$) & $1.4044 \times 10^{12}$ \\
Log-likelihood                  & -3373.02 \\
AIC                             & 6770.03 \\
AICc                            & 6771.55 \\
BIC                             & 6810.64 \\
\bottomrule
\end{tabular}
\end{table}

\subsection{General Interpretation}

\quad \quad \, \textbf{Non-seasonal Components}:  
The AR and MA coefficients indicate how past values of the series and past errors influence the current value. Although the AR coefficients (ar1, ar2, ar3) have relatively small magnitudes, the negative MA component (ma1) suggests a corrective effect on forecast errors.

\medskip

\textbf{Seasonal Components}:  
The seasonal coefficients (SAR and SMA) highlight the influence of periodic patterns, particularly in monthly series. The signs and magnitudes of the coefficients indicate that the series is strongly impacted by effects that repeat annually.

\medskip

\textbf{Fit Quality}:  
The AIC, AICc, and BIC indicators, along with the log-likelihood, provide parameters to compare this model with alternatives. Lower values in these indicators suggest a good balance between data fit and model complexity.

\end{document}